\documentclass[11pt]{article}
\usepackage{amsmath, amssymb}
\begin{document}
%
\newcommand{\RR}{\mathbb{R}} 
\newcommand{\Eins}{{\mathbf 1}} 
\newcommand{\e}{{\rm e}}
\renewcommand{\t}{{\rm t}}
\newcommand{\inv}{^{-1}}
\newcommand{\Mat}{\hbox{\rm Mat}}
\renewcommand{\theequation}{\thesection.\arabic{equation}}
\title{\bf A comment on the dual field \\ in the AdS-CFT correspondence}
\author{{\sc M. D\"utsch\thanks{E-mail: 
        {\tt duetsch@theorie.physik.uni-goe.de}}} {} and 
        {\sc K.-H. Rehren\thanks{E-mail:
        {\tt rehren@theorie.physik.uni-goe.de}}} \\ 
Institut f\"ur Theoretische Physik, Universit\"at G\"ottingen, \\
37073 G\"ottingen,  Germany }
\maketitle

\begin{abstract} In the perturbative AdS-CFT correspondence, the dual 
  field whose source are the prescribed boundary values of a bulk field in 
  the functional integral, and the boundary limit of the quantized bulk 
  field are the same thing. This statement is due to the fact that Witten 
  graphs are boundary limits of the corresponding Feynman graphs for the 
  bulk fields, and hence the dual conformal correlation functions are limits 
  of bulk correlation functions. This manifestation of duality is
  analyzed in terms of the underlying functional integrals of
  different structure. 
\end{abstract}

PACS 2001: 11.10.-z, 11.25.-w

MSC 2000: 81T20, 81T30

\section{Introduction}
The AdS-CFT correspondence \cite{GKP,W} as a concretization of Maldacena's 
conjecture \cite{M} owes much of its fascination to the fact that it 
produces conformal correlation functions for the ``dual'' conformal 
field $\mathcal O$, using as the generating functional a functional
integral of highly non field theoretical appearance, of the form
\begin{equation}
\left\langle\e^{\mathcal{O}(f)}\right\rangle = 
\frac{Z(f)}{Z(0)} \quad\hbox{with}\quad 
Z(f) = \int D\phi\;\e^{-I(\phi)} \; \delta(\phi_0-f),
\end{equation}
where $\phi_0$ is the boundary limit of the functional variable $\phi$,
and $\mathcal{O}$ is the (Euclidean) dual field. The use of {\em prescribed
values} of the functional integration variable $\phi$ as the {\em
source} for a quantum field, is deeply inspired from string theory
\cite{GKP,W} and has no precedent in field theory. We shall refer to
(1.1) as the ``dual prescription''. 

It was noticed soon (e.g., \cite{BDHM}) that the bulk-to-boundary
propagators derived from the the dual prescription are limits of the
bulk-to-bulk propagators as one of the bulk coordinates approaches the
boundary. This implies that the dual Green functions are boundary
limits of bulk Green functions, and hence the dual conformal fields
themselves are boundary limits, or ``restrictions'', of the bulk
fields in the sense of eqs.\ (1.7), (4.1) below.%
\footnote{The boundary limit of the bulk field is a special case of
  the restriction of a quantum field to a time-like hypersurface, 
  which is well defined \cite{Bo} and yields (non-Lagrangian) 
  quantum fields in one dimension less. In contrast, restrictions of 
  quantum fields to space-like surfaces (``time zero fields'') \cite{H} 
  or light fronts \cite{Sch} are generally too singular to exist.}

The recognition of the dual field as a boundary limit of the
bulk field perfectly complies with two results on the AdS-CFT 
correspondence derived in axiomatic frameworks. 
In the Wightman axiomatic framework is has been shown \cite{BBMS} that 
the boundary limits of AdS correlation functions inherit the properties 
of locality, covariance, energy positivity and Hilbert space positivity 
(unitarity), and fulfill the physical requirements of a local conformal 
QFT in Minkowski space. A similar conclusion can be drawn from the
algebraic treatment in \cite{KHR} where the local observables of a QFT
on AdS and a corresponding conformal QFT are identified, such that the
sharply localized conformal observables coincide with 
the AdS observables close to the boundary. (The algebraic treatment
also allows to characterize and define the observables in the interior
of AdS in terms of conformal observables.)  

Indeed, many tests of the field theoretic properties of correlation 
functions computed with the dual prescription have produced perfectly 
sensible results (including operator product expansions, 
Ward identities, and positivity \cite{FMMR,HPR,OK}).

The generating functional for the conformal boundary correlations 
should therefore be as well representable as a functional integral where 
the field is coupled in the usual field theoretic way to a source, with the 
specification that the source is confined to the boundary,
\begin{equation}
\left\langle\e^{\phi_0(f)}\right\rangle = 
\frac{\widetilde Z(f)}{\widetilde Z(0)} \quad\hbox{with}\quad 
\widetilde Z(f) = \int D\phi\;\e^{-I(\phi)} \;
\e^{ \int\phi_0f}.
\end{equation}
On the right-hand side of this formula, $\phi_0$ stands for the boundary 
limit of the integration variable $\phi$, while on the left-hand side it 
denotes the Euclidean bulk quantum field, restricted to the boundary. 

It is the purpose of this letter to understand the way how the two competing 
functional integrals (1.1) and (1.2) of drastically different appearance can 
provide the same results. $\widetilde Z(if)$ is actually the functional 
Fourier transform of $Z(f)$. Coincidence of the Schwinger functions generated 
by the two functional integrals seems to imply that the integrals coincide 
and hence must be essentially their own functional Fourier transforms,
for any form of the action $I(\phi)$. This looks like a straight absurdity.

The apparent conflict is resolved by the fact that we shall have to 
specify the function spaces to which the respective functional measures 
$D\phi$ apply, which is equivalent to the choice of the propagators $G_+$,
$G_-$, formally giving rise to two different functional integrals 
$Z^\pm(f)$, $\widetilde Z^\pm(f)$ of either type (1.1), (1.2), the 
superscript distinguishing the two measures. Then we observe that 
in the dual case, the implementation of the $\delta$-function
in (1.1) leads to an effective modification of the propagator. The
total dual bulk-to-bulk propagator $\Gamma_-$ turns out to coincide 
with the field theoretic propagator $G_+$,
\begin{equation}
\Gamma_-(z,x;z',x') = G_+(z,x;z',x').
\end{equation}
Likewise, we shall analyze the consequences of the dual prescription 
for the bulk-to-boundary propagator $K_-$ and for the tree level
2-point function (the only connected graph without vertices), with the 
result that 
\begin{equation}
K_-(z,x;x') = c\cdot \lim_{z'\to 0} z'{}^{\Delta_+} G_+(z,x;z',x')
\end{equation}
where $\Delta_+$ is the scaling dimension of the boundary field and $c$
a numerical coefficient, and the tree level 2-point function equals 
\begin{equation}
c^2\cdot \lim_{z'\to 0} z'{}^{\Delta_+} \lim_{z\to 0} z^{\Delta_+} 
G_+(z,x;z',x').
\end{equation}
The right-hand sides of eqs.\ (1.4), (1.5) are the appropriate limits of the 
field theoretical propagator when the source is confined to the boundary.
Hence eqs.\ (1.3--5) imply 
\begin{equation}
Z^-(f) = \widetilde  Z^+(c\cdot f),
\end{equation} 
valid graph by graph in the formal Euclidean 
perturbation series. This in turn implies that the conformal field
$\mathcal{O}^-$ defined by the dual prescription (1.1) coincides with
$\phi_0^+$ defined by the restriction prescription (1.2),
\begin{equation}
\mathcal{O}^-(x) = c\cdot\phi^+_0(x) \equiv 
c\cdot \lim_{z\to 0}z^{-\Delta_+}\phi^+(z,x). 
\end{equation}

We believe that these facts, which we systematically establish for the most 
general scalar and vector fields, are a non-trivial manifestation of duality 
in the AdS-CFT correspondence. They pertain to the approximation of the 
holographic AdS-CFT correspondence in which string effects are suppressed and 
gravity is treated as a tensor field in a fixed background. Indeed, the 
action may be any local functional involving a finite number of tensor
fields. Supersymmetry or ``large $N$'' is not assumed.

We emphasize that these results concern the formal perturbative 
expansions of the Euclidean boundary field theories in question,
subject to the well-known difficulties encountered in the Euclidean 
functional integral approach. Clearly, individual graphs require 
renormalization, and the entire series diverges. Moreover, the
correlation functions may fail to satisfy the Osterwalder-Schrader (OS) 
positivity condition \cite{OS}, which is crucial in order to qualify as 
Schwinger functions of an associated real-time QFT. Only OS positivity 
guarantees Einstein causality, Hilbert space positivity and positivity 
of the energy. The positivity property of the functional integral 
inherited from the Gaussian measure is not sufficient in this respect.  

The graph-by-graph identification (1.6) is not affected by renormalization 
(if the same renormalization conditions are imposed) and analytic
continuation. One may therefore expect that a proper renormalized
real-time interpretation of (1.1) also coincides with the real-time
perturbation theory for a bulk field with subsequent restriction.

\section{A discrete model}
\setcounter{equation}{0}
We want to emphasize the basically algebraic nature of the relations
among the relevant propagators, pertaining to the passage between
source terms of the respective forms $\e^{\phi_0\cdot f}$ and
$\delta(\phi_0-f)$. For this purpose we first consider
finite-dimensional Gaussian integrals, replacing anti-deSitter space
by a lattice. In the finite-dimensional case, formal manipulations
with Gaussian integrals are exact. In particular, there is no room for
further specifications of propagators, and the generating functionals
$Z(if)$ and $\widetilde Z(f)$ are definitely distinct. We shall see
that the difference resides entirely in the propagators.  
Their algebraic characterizations established in this section will be
exploited in the next sections for the continuum case.  

We replace the real functions $\phi_\alpha(x)$ by an $N$-tuple of
integration variables $\phi \equiv (\phi_i)_{i=1\dots N}\in\RR^N$
where the index $i$ labels both the lattice points and the Lorentz
(multi)indices $\alpha$ in the case of a tensor field. We arrange the
numbering such that $i=1\dots n$ label the boundary variables (boundary 
values of the field), while the remaining ones label the bulk variables. 

We denote by $e:\RR^n\to\RR^N$ the corresponding embedding of
the spaces of integration variables, and by $e^\t:\RR^N\to\RR^n$ its
adjoint. The boundary variables are thus $\phi_0\equiv e^\t\phi \in\RR^n$. 

The quadratic part $\frac12 (\phi,A\phi)$ of the action is given by a
symmetric matrix $A \in \Mat_N(\RR)$. The total action is of the form 
\begin{equation} \textstyle
I(\phi) = \frac12 (\phi,A\phi) + V(\phi)
\end{equation}
with a local polynomial potential $V(\phi)=\sum_{i}v(\phi_{i})$.
We proceed in the usual perturbative way by expanding $\exp-V(\phi)$ as a
power series, and performing the Gaussian integrals. 

The integral $\widetilde Z(f)$, $f\in\RR^n$, involving the source term
$\exp (\phi_0,f) \equiv \exp (e^\t\phi,f) \equiv \exp (\phi,ef)$ is
computed as usual by completing the square and shifting the
integration variable $\phi \to \phi + A\inv ef$. This yields
\begin{equation}
\widetilde Z(f) = \e^{\frac12(f,\alpha f)}\cdot
\int D\phi\;\e^{-\frac12(\phi,A\phi)}\exp-V(\phi+A\inv ef).
\end{equation}
The Gaussian prefactor comes from $(ef,A\inv ef) = (f,\alpha f)$,
where $\alpha$ is the $n \times n$ matrix 
\begin{equation}
\alpha:=e^\t A\inv e \in \Mat_n(\RR).
\end{equation}

For the integral $Z(f)$, $f\in\RR^n$, with the source term
$\delta(\phi_0-f)$ we use the projections $E=ee^\t$ (``boundary'') and
its complement $E^\perp=\Eins_{N}-E$ (``bulk'') to separate the
boundary variables from the bulk variables:  
\begin{equation}
\phi=E\phi+E^\perp\phi\equiv e\phi_0 + E^\perp\phi,
\end{equation} 
and perform the obvious integration over the boundary variables
$\phi_0$, thus $\phi=e f + E^\perp\phi$. In order to decouple
the external variables $f$ from the integration variables $E^\perp\phi$, 
we shift the latter by $E^\perp A\inv (e \alpha\inv f)$ such that 
$\phi= E^\perp\phi'+e f+E^\perp A\inv (e \alpha\inv f)$. 
Writing $e f = EA\inv(e \alpha\inv f)$, we get 
$\phi=E^\perp\phi' + A\inv (e \alpha\inv f)$. 
The quadratic term decouples as desired:
\begin{equation}
(\phi, A\phi) = (f,\alpha\inv f)+(E^\perp\phi',AE^\perp\phi'),
\end{equation}
and the functional integral becomes (suppressing the prime)
\begin{align}
Z(f) = \e^{-\frac12(f,\alpha\inv f)} & \nonumber \\
\times \int D(E^\perp\phi)\; & \e^{-\frac12(E^\perp\phi,AE^\perp\phi)}
\exp-V(E^\perp\phi + A\inv (e \alpha\inv f)).
\end{align}

From these formulae (2.2), (2.6), we read off the diagrammatical rules. 
The vertices, given by the polynomial structure of the potential $v$,
are common to both functional integrals. They involve a summation over
the lattice index $i$. Due to the respective shifts of the variable
$\phi_{i}$, there are inner lines (corresponding to the integration
variables) and outer lines (corresponding to the external variables
$f$) attached to each vertex. 

The ``bulk-to-bulk propagator'' for the inner lines connecting two vertices 
is the inverse of the Gaussian covariance matrix of the respective integral. 
The ``bulk-to-boundary'' propagator for the outer lines is the ($N\times n$ 
matrix-valued) coefficient of $f$ in the shifted argument of $V$.  

For $\widetilde Z(f)$ with the exponential insertion, eq.\ (2.2), we
read off the bulk-to-bulk propagator
\begin{equation}
G=A\inv
\end{equation}
and the bulk-to-boundary propagator which is the right boundary
restriction of $G$, while $\alpha$ giving the leading Gaussian is its
two-sided restriction, 
\begin{equation}
H = G e, \qquad \alpha = e^\t Ge = e^\t H.
\end{equation}
 
For $Z(f)$ with the $\delta$ function insertion, eq.\ (2.6), the
bulk-to-bulk propagator is obtained as follows. Since only the bulk
variables $E^\perp\phi$ propagate, $\Gamma$ should have vanishing
$\RR^n$ (boundary) components. On the orthogonal (bulk) subspace
$E^\perp\RR^N$, $\Gamma$ should be the inverse of $A$. Hence 
\begin{equation}
E\Gamma=0=\Gamma E \quad\hbox{\rm and}\quad E^\perp
A\Gamma=E^\perp=\Gamma AE^\perp. 
\end{equation}
This pair of algebraic conditions determines the matrix $\Gamma$ uniquely as 
\begin{equation}
\Gamma = G - G e \, \alpha\inv \, e^\t G.
\end{equation}
The bulk-to-boundary propagator is
\begin{equation}
K=G e \, \alpha\inv \equiv H\alpha\inv,
\end{equation}
and can be uniquely characterized by the pair of algebraic conditions 
\begin{equation}
EK=e \; \Leftrightarrow \; e^\t K=\Eins_n \quad\hbox{\rm and}\quad 
E^\perp AK=0. 
\end{equation}

We recognize in (2.9) a discrete version of Dirichlet boundary
conditions for the inverse of $A$ on the bulk. This property will be 
crucial when we pass to the continuum in the next section.%
\footnote{It is also
  instructive to pass to the other extreme in which the lattice
  consists of only two points $1$ (the boundary) and $2$ (the bulk),
  i.e., $n=N-n=1$. In this case, with $A=\left(\begin{array}{cc}a&b\cr
  b&c \end{array}\right)$ and $A\inv= \left(\begin{array}{cc}\alpha
  &\beta\cr\beta &\gamma\end{array}\right)$, one has 
  $\Gamma= \left(\begin{array}{cc}0&0\cr 0&1/c \end{array}\right)$ where
  $1/c=\gamma-\beta^2/\alpha$.}
 
We conclude that the two functional integrals $Z(f)$ and $\widetilde Z(f)$
are obtained as sums over the same sets of graphs but with different
prescriptions for the propagators to be inserted for the internal and
external lines, and with different leading Gaussian prefactors.

\section{Scalar propagators on AdS}
\setcounter{equation}{0}
Passing to Euclidean field theory on $d+1$-dimensional anti-deSitter
space, we substitute the real scalar field $\phi(z,x)$ (in the usual
coordinates $z\in\RR_+$, $x\in\RR^d$) for the vector
$\phi$, and the Klein-Gordon operator for the matrix $A$: 
\begin{equation}
A=-\square_g+M^2=
-z^{1+d}\partial_zz^{1-d}\partial_z - z^2{\sum}_{i=1}^d\partial_i^2 + M^2. 
\end{equation}
The inner product
$(\phi,A\phi)$ is the bulk integration with measure
$dz\,d^dx\,\sqrt g$, $\sqrt g=z^{-1-d}$. The potential has the form
$V(\phi)=\int dz\,d^dx\,\sqrt g\, v(\phi(z,x))$ with some polynomial
density $v(\phi)$.

The inverse $G = A\inv$ is the Green function solving 
\begin{equation}
(-\square_g + M^2)G(z,x;z',x')=z^{1+d}\delta(z-z')\delta^d(x-x').
\end{equation}
There are two linearly independent AdS-invariant solutions,
\begin{equation} \textstyle
G_\pm(z,x;z',x') = \gamma_\pm \cdot (2u)^{-\Delta_\pm}
{}_2F_1(\Delta_\pm,\Delta_\pm+\frac {1-d}2,2\Delta_\pm+1-d;-2u\inv) 
\end{equation}
where $u=\frac{(z-z')^2+(x-x')^2}{2zz'}$,
$\Delta_\pm$ are the two solutions of $\Delta(\Delta-d)=M^2$,
\begin{equation} \textstyle
\Delta_\pm = \frac d2 \pm \frac12\sqrt{d^2+4M^2},
\end{equation} 
and the normalization coefficients are $\gamma_\pm= \frac{\pi^{-\frac d2}
\Gamma(\Delta_\pm)} {2\Gamma(\Delta_\pm+1-\frac d2)}$. 
The two solutions are distinguished by the boundary behaviour   
\begin{equation}
G_\pm \sim z^{\Delta_\pm}\quad{\rm as}\quad z\to 0
\end{equation} 
(and likewise for $z'$). The choice of either of them therefore
specifies the functional integration measure to extend formally over
spaces of functions $\phi^\pm(z,x)$ with the corresponding boundary
behaviour $\sim z^{\Delta_\pm}$.%
\footnote{The Klein-Gordon operator (3.1) is homogeneous in $z$ near
  the boundary and hence preserves spaces of functions $\phi$ which
  behave like $\sim z^\Delta$ near $z=0$. The integral 
  $(\phi,A\phi)= \int z^{-1-d}dz\,dx\; \phi\,(-\square_g+M^2)\phi$
  converges at $z=0$ and is symmetric as a quadratic form only in the
  case of $\Delta_+ > \frac d2$. 
  Nevertheless, proceeding formally also in the case of
  $Z^-$ where $\Delta_- < \frac d2$, will turn out to be
  justified (due to the suppression of the boundary functional
  integration variables by the $\delta$ function), and in fact match the
  perturbative rules adopted in the literature \cite{FMMR,GKP,KW,W}.} 
We denote the corresponding integrals (1.1) and
(1.2) by $Z^\pm(f)$ and $\widetilde Z^\pm(f)$.

We retain from the discrete model the diagrammatical rules. The field
theoretical integrals $\widetilde Z^\pm(f)$ 
have Gaussian prefactors $\exp{\frac12(f,\alpha_\pm f)}$
and involve propagators $G_\pm$ (bulk-to-bulk) and $H_\pm$
(bulk-to-boundary), while the AdS-CFT dual integrals $Z^\pm(f)$ 
have prefactors $\exp{-\frac12(f,\alpha_\pm\inv f)}$ and involve 
propagators $\Gamma_\pm$ and $K_\pm$. The various propagators are
obtained from $G_\pm$ via the algebraic relations (2.3) and
(2.7)--(2.12), to be understood as relations among integration kernels
$G_\pm$, $\Gamma_\pm$, $H_\pm$, $K_\pm$, $\alpha_\pm$ (some of which
will require regularization) rather than matrices, and $e_\pm$ stand
for rescaled limits of the form   
\begin{equation}
(F_\pm e_\pm)(x) := \lim_{z\to 0}z^{-\Delta_\pm}F_\pm(z,x).
\end{equation}
The graphs to be summed in both integrals are the same, with the same
vertices, but different propagators. 

For the field theoretical integrals $\widetilde Z^\pm(f)$, the 
bulk-to-bulk propagators are $G_\pm(z,x;z',x')$ as in eq.\ (3.3). 
The bulk-to-boundary propagators are, according to eq.\ (2.8), 
the limits 
\begin{equation}
H_\pm(z,x;x') = \lim_{z'\to 0} z'{}^{-\Delta_\pm} G_\pm(z,x;z',x')
 =\gamma_\pm\cdot \left(\frac{z}{z^2+(x-x')^2}\right)^{\Delta_\pm}
\end{equation}
and likewise, according to eq.\ (2.3), the tree level 2-point
functions are
\begin{equation}
\alpha_\pm(x,x') = \lim_{z\to 0}
{z}^{-\Delta_\pm}H_\pm(z,x;,x') = \gamma_\pm\cdot (x-x')^{-2\Delta_\pm}.
\end{equation}
Thus, the boundary fields $\phi_0^\pm$ have the scaling dimensions
$\Delta_\pm$ (at tree level).

To compute the propagators $\Gamma\equiv G - H\alpha\inv H^\t$ and
$K=H\alpha\inv$ for the dual integrals $Z^\pm(f)$ according to 
(2.10) and (2.11), would involve the determination of, and
multiplication with inverse integral kernels $\alpha_\pm\inv$. 
It turns out advantageous to exploit instead the algebraic
characterizations (2.9)  
and (2.12) of the dual propagators, worked out in Sect.\ 2. 

Translated into the continuum context, (2.9) states that $\Gamma_\pm$
solve Green's differential equation in the bulk, and vanish on the
boundary. In other words, they are the Green functions with Dirichlet
conditions with respect to the restrictions given by the limits
$e_\pm$. Now, by (3.5) and $\Delta_+ > \Delta_-$, the Green function
$G_+$ vanishes faster than $G_-$ and hence satisfies the Dirichlet
condition with respect to the limit $e_-$. We conclude that 
\begin{equation}
\Gamma_-  = G_+.
\end{equation}
Likewise, (2.12) translates into the conditions that $K_\pm$ solve the
Klein-Gordon equation in the bulk, and approach $\delta^d(x-x')$ in
the limits $e_\pm$. By definition, the first condition for $K_\pm$ is
fulfilled by $H_\mp$. By virtue of a simple scaling argument \cite{W}
based on the relation   
\begin{equation}
\Delta_+ + \Delta_- = d,
\end{equation}
$H_+$ also fulfills the second condition for $K_-$ up to a normalization,
\begin{equation}
(e_-^\t H_+)(z,x;x') \equiv \lim_{z\to 0} z^{-\Delta_-} H_+(z,x;x') =
c\inv\cdot \delta(x-x').  
\end{equation}
The constant is computed as $c = 2\Delta_+ - d =: c(\Delta_+)$. Hence 
\begin{equation}
K_- = c \cdot H_+. 
\end{equation}
By (2.11), this implies the integral identity
\begin{equation}
c \cdot H_+\alpha_- = H_-
\end{equation}
involving a perfectly regular $\RR^d$ integration. In contrast,
replacing $\Delta_\pm$ by $\Delta_\mp$ everywhere, $H_-\alpha_+$ is
UV-divergent. We UV-regularize $\alpha_+$ by analytic continuation of
(3.13),%
\footnote{Regarding $H_\pm$, $\alpha_\pm$ as functions of $\Delta_\pm$,
   and $\Delta_-$ as a function of $\Delta_+$, (3.13) is an equality
   $c(\Delta)\cdot H(\Delta) \alpha(d-\Delta) = H(d-\Delta)$ of two
   meromorphic functions, valid at $\hbox{Re }\Delta > \frac d2$ 
   (hence at $\Delta=\Delta_+$). The right-hand side being
   analytic also at $\hbox{Re }\Delta < \frac d2+1$, it defines the
   analytic continuation of the left-hand side to $\Delta = \Delta_-$. 
   At this point, $H(\Delta_-)=H_-$, $H(d-\Delta_-)=H_+$,
   $\alpha(d-\Delta_-)=\alpha_+$, and $c(\Delta_-)=-c$.}
such that also
\begin{equation}
-c\cdot H_-\alpha_+ = H_+
\end{equation}
holds. Applying the limit $e_-$ to both sides of (3.14), using 
$e_-^\t H_- = \alpha_-$ and $c\cdot e_-^\t H_+ = \Eins$, we get
$-c^2\cdot \alpha_-\alpha_+ = \Eins$ or  
\begin{equation}
\alpha_-\inv = -c^2\cdot\alpha_+.
\end{equation}
As $-\alpha_\pm\inv(x,x')$ are the tree
level 2-point functions of the dual fields $\mathcal{O}^\pm$, the latter
have the scaling dimensions $\Delta_\mp$. The regularization of
$\alpha_+$ implicit in (3.14) is in agreement with the one adopted in
\cite{KW}, and the absolute normalization $c^2\gamma_\mp$ of the
2-point functions of $\mathcal{O}^\pm$ inferred from (3.15) is in
agreement with the correction advocated in \cite{FMMR,KW}.     

Now, by virtue of the identifications (3.9), (3.12) and (3.15), the
propagators involved in $Z^-(f)$ and $\widetilde Z^+(f)$ are the same,
up to the numerical factors. This proves the assertion (1.6), and
hence (1.7), with $c =\sqrt{d^2+4M^2}$. 

Scrutinizing the above argument, 
we observe that most of it follows from the
algebraic characterizations of the propagators obtained in Sect.\ 2. 
The only independent information was the validity of the limit (3.11) 
which in turn followed by a scaling argument from the relation (3.10),
along with the analytic property of the coefficient as a function of
$\Delta_+$ 
\begin{equation}
c(d-\Delta_+) = -c(\Delta_+)
\end{equation} 
ensuring the correct relative normalizations of the coefficients
in (1.4), (1.5). 

\section{Vector fields}
\setcounter{equation}{0}
We want to generalize the previous argument to vector fields
$\phi_\mu(z,x)$, $\mu=z,0,\dots d-1$. For vector fields, the 
restriction maps $e_\pm$ involve rescaled limits {\em and} the
projection onto the transverse components $\phi_i$, $i=0,\dots d-1$. 
We shall establish the identity 
\begin{equation}
\mathcal{O}^-_i(x) = c\cdot (\phi^+_0)_i(x) \equiv
c \cdot \lim_{z\to 0}z^{1-\Delta_+} \phi^+_i(z,x).
\end{equation}
The dimensions $\Delta_\pm$ will be determined from the quadratic
part of the action, and satisfy again (3.10). 
Adapting the remark at the end of the previous section, we observe
that we only have to compute the coefficient $c$ in
\begin{equation}
\lim_{z\to 0} z^{1-\Delta_-} \big(\lim_{z'\to 0}z'{}^{1-\Delta_+} 
G_{+,i j}(z,x;z',x') \big) = c\inv\cdot \delta_{ij}\delta(x-x'),
\end{equation}
as a function of $\Delta_+$ and verify that it again satisfies eq.\ (3.16).

We shall only sketch the computation. The most general quadratic action 
\begin{equation} 
\int dz\,d^dx\,\sqrt g \;\textstyle \bigg(\frac14 F_{\mu\nu}F^{\mu\nu} +
\frac12\lambda (D^\mu\phi_\mu)^2 + \frac12 M^2 \phi^\mu\phi_\mu\bigg)
\end{equation}
where $F_{\mu\nu} = D_\mu\phi_\nu-D_\nu\phi_\mu$, gives rise to Green's
differential equation  
\begin{equation}
{A_\mu}^\nu G_{\nu \alpha}(z,x;z',x') =
g_{\mu\alpha}z^{1+d}\delta(z-z')\delta^d(x-x') 
\end{equation}
with the differential operator
\begin{equation}
A_{\mu\nu}= (-D_\kappa D^\kappa+M^2-d)g_{\mu\nu} + (1-\lambda)D_\mu D_\nu
\end{equation}
(the shift in the mass being due to the curvature). We make an ansatz
with the most general AdS-covariant bivector (with
$u=\frac{(z-z')^2+(x-x')^2}{2zz'}$ as before)
\begin{equation}
G_{\mu \alpha}(z,x;z',x') = -g_1(u) \cdot (\partial_\mu\partial'_\alpha u) -
g_2(u) \cdot (\partial_\mu u)(\partial'_\alpha u).
\end{equation}
The singularity in (4.4) requires the short-distance behaviour of the
functions $g_1\approx\gamma_1'u^{-\frac{d-1}2}$ and 
$g_2\approx\gamma_2'u^{-\frac{d+1}2}$ at $u\approx 0$, and
\begin{equation} \textstyle 
(d-1)\gamma'_1 + 2 \gamma'_2 = (2\pi)^{-\frac{d+1}2}\Gamma(\frac{d+1}2).
\end{equation}
At $u\neq 0$, (4.4) yields two differential equations of second
order, which can be decoupled with the help of $f=g_2-g_1'$: 
\begin{eqnarray}
u(u+2)f'' + (d+3)(u+1)f' + (2d-M^2)f = 0, \\[2mm]
u(u+2)g_1'' + (d+3)(u+1)g_1' + (d+1-M^2/\lambda)g_1 = \qquad \nonumber
\\ \textstyle 
= \frac{1-\lambda}\lambda u(u+2)f' + (\frac{1-\lambda}\lambda d-2)(u+1)f.
\end{eqnarray}
The homogeneous equation (4.7) for $f(u)$ is solved by
\begin{equation} \textstyle 
f_\pm(u) = \gamma_\pm\cdot u^{-\Delta_\pm-1}
{}_2F_1(\Delta_\pm+1,\Delta_\pm+\frac{1-d}2,2\Delta_\pm+1-d;-2u\inv)
\end{equation}
with
\begin{equation} \textstyle 
\Delta_\pm = \frac d2 \pm \frac 12 \sqrt{(d-2)^2+4M^2}.
\end{equation}
Both solutions have the same short-distance behavior 
$f\approx\gamma'u^{-\frac{d+1}2}$. Because of $f=g_2-g_1'$, the
coefficient is $\gamma'=\gamma_2'+\frac{d-1}2\gamma_1'$, which is determined by
(4.7). This fixes the absolute normalizations in (4.10),
\begin{equation} \textstyle
\gamma_\pm = 2^{-\Delta_\pm-1}\pi^{-\frac d2}
\frac{\Gamma(\Delta_\pm+1)}{\Gamma(\Delta_\pm+1-\frac d2)}.
\end{equation}

We choose either $f_+$ or $f_-$ and suppress the subscript for the moment.
The function $g_1\equiv g_{1\pm}$ is determined by the inhomogeneous
equation (4.9) up to a solution of the corresponding homogeneous
equation. One of these is too singular at short distance and can be
excluded. The less singular solution 
\begin{equation} \textstyle 
{}_2F_1(q_+,q_-,\frac{d+3}2;-u/2), \qquad
q_\pm = \frac{d+2}2 \pm \frac 12\sqrt{d^2+4M^2/\lambda}
\end{equation}
behaves like $u^{-q_-}$ at large $u$ (small $z$, $z'$). If this
solution would dominate the behavior of $g_1$ at large $u$ (in case
$q_-<\Delta$), then $g_1'$ would dominate $f$, and $g_2 \approx g_1'$. 
This would produce a boundary 2-point function of scaling dimension
$q_-$, which violates conformal invariance, however. Thus we must seek
the special solution $g_1$ whose behavior at large $u$ is determined
not by (4.13), but by the inhomogeneity of (4.9). For this solution,
$g_1\approx\gamma_1u^{-\Delta}$ and $g_2 = g_1'-f
\approx\gamma_2u^{-\Delta-1}$, with 
\begin{equation}  \textstyle 
\gamma_1 = \frac\gamma{\Delta-1} = \frac {2^{-\Delta-1}\pi^{-\frac d2}}
{\Delta-1}\frac{\Gamma(\Delta+1)}{\Gamma(\Delta+1-\frac d2)}, \qquad
\gamma_2=-\Delta\gamma_1-\gamma = -\gamma_1. 
\end{equation}
This controls the boundary behavior of the Green function. 
With
\begin{eqnarray}
H_{\mu j}(z,x;x') = \lim_{z'\to 0} z'{}^{1-\Delta} G_{\mu j}(z,x;z',x')
= \qquad\qquad\qquad\qquad\qquad\quad \\ = \textstyle \gamma_1 
\left(\frac{2z}{z^2+(x-x')^2}\right)^\Delta\partial_\mu\frac{(x-x')_j}z 
+ \gamma_2 \left(\frac{2z}{z^2+(x-x')^2}\right)^{\Delta+1} 
\frac{(x-x')_j}z\partial_\mu\frac{z^2+(x-x')^2}{2z}, \nonumber
\end{eqnarray}
we obtain the conformally invariant tree level 2-point function
\begin{eqnarray}
\alpha_{i j}(x;x') = \lim_{z\to 0}z^{1-\Delta}H_{i j}(z,x;x') =2^\Delta\gamma_1  
\cdot\frac{\delta_{ij}-2\frac{(x-x')_i(x-x')_j}{(x-x')^2}}{(x-x')^{2\Delta}}. 
\end{eqnarray}
In particular, $\Delta\equiv\Delta_\pm$ given in (4.11) are the
scaling dimensions of the boundary fields, satisfying (3.10) as
announced.  

We can now also compute the limit $e_-^\t H_+$ and find 
\begin{equation}
\lim_{z\to 0} z^{1-\Delta_-} H_{+,i j}(z,x;x') = (2\Delta_+-d)\inv\cdot
\delta_{ij}\delta^d(x-x').
\end{equation}
Thus, $c = 2\Delta_+ -d$ is the same analytic function of $\Delta_+$
as for scalar fields, satisfying (3.16) as desired. 

As explained above, this completes the proof for the validity of (1.6)
also for vector fields. More explicitly, eq.\ (4.1) holds with
$c= \sqrt{(d-2)^2+4M^2}$. The case of massless gauge fields corresponds to
$M^2=0$, $\Delta_+=d-1$ (independent of the gauge parameter
$\lambda$), so that (4.16) is the 2-point function of a conserved
current. For tensor fields of higher spin, the most general
bi-covariant ansatz for the Green functions involves more unknown
functions, which complicates the analysis. But since only (3.10) and
(3.16) need to be verified, the comparison of (3.4) with (4.10) and
the coincidence of the function $c(\Delta_+)$ both for scalar and
vector fields, lend support to the expectation that the result
generalizes to tensor fields of arbitrary rank.    

\section{Conclusion}
\setcounter{equation}{0}
The perturbative expansion of the dual field $\mathcal{O}^- $ in terms 
of ``Witten graphs'' matches the canonical (field theoretical) expansion 
of the interacting field $\phi^+$ in the bulk of AdS, with subsequent 
restriction to the boundary. We have presented a structural analysis
of this fact (which was previously observed, e.g., by \cite{BDHM}) in
terms of the formal identification (1.6), graph by graph, of the
generating functionals for the respective Euclidean correlation
functions. The relations between the relevant propagators, 
pertaining to the passage between source terms of the respective forms
$\e^{\phi_0\cdot f}$ and $\delta(\phi_0-f)$, are basically algebraic,
and independent of any specific geometry. The solution (1.3--5) to
these relations, in contrast, is largely due to $SO(1,d+1)$ symmetry,
while the only piece of the argument which seems not automatic, is the
analytic property (3.16) of the coefficient function $c(\Delta_+)$
appearing in (3.11).     

In the free case, $V(\phi)=0$, both integrals (1.2) and (1.1) give
rise to purely Gaussian boundary fields with scaling dimensions 
$\Delta_\pm=\frac d2\pm\frac 12\sqrt{d^2+4M^2}$. Their Euclidean 
correlations satisfy OS positivity provided the dimension satisfies the
unitarity bound $\Delta\geq \frac d2-1$. Thus, $\phi^+_0$ is
always related to a real-time quantum field, and so is $\phi^-_0$
provided $\frac 12\sqrt{d^2+4M^2}\leq 1$. 

These are Gaussian fields with non-canonical dimension $\Delta$. Such
fields belong to the class of ``generalized free fields'' \cite[Ch.\
2.6]{J} which were first introduced in \cite{L} as asymptotic fields
appropriate when a particle interpretation breaks down (e.g., in
conformal theories). The $n$-point functions of a generalized free
field factorize into 2-point functions, and its commutator is a
numerical distribution, but there is no Lagrangian description with an
equation of motion because the K\"allen-Lehmann measure $\rho(m^2)$  
$$
\langle\Omega,\varphi(x)\varphi(y)\Omega\rangle = \int_0^\infty dm^2
\rho(m^2) \Delta_m^+(x-y)
$$
may cover a continuum of masses. (Specifically, for $\phi_0^\pm$, 
$\rho(m^2)\sim m^{2\Delta_\pm-d}$.) Thus a generalized free field can
have the same 2-point function as an interacting field which
necessarily covers a continuum of masses extending to $\infty$
\cite[Ch.\ 6.1]{J}. It will be shown elsewhere \cite{DR} that inspite
of the absence of an equation of motion, a stress-energy tensor can be
defined for generalized free fields, which is more singular than a
Wightman field but still is a local density for the generators of
space-time symmetries.    

The (real time) AdS-CFT correspondence thus amounts to a perturbation
around a conformal generalized free field whose non-canonical dimension 
is not itself a perturbative effect, unlike an anomalous dimension. 

We notice that a standard perturbation theory around generalized
free fields has not been formulated so far, and is expected to suffer
from aggravated renormalization problems: e.g., in the case of
$\phi^+_0$ already the integration for the retarded propagator 
$$
[\phi^+_0(x),\phi^+_0(x')]\theta(x^0-x'{}^0) = \int_0^\infty
dm^2m^{2\nu} \Delta_m^{\rm ret}(x-x') 
$$
is UV divergent. Thus, the {\em free} propagator itself requires
renormalization, i.e., its distributional extension to the diagonal
$x=x'$ is non-unique \cite{BF2}. Moreover, the propagator entering the
power counting argument with a larger scaling dimension affects
renormalizability always for the worse. 

But perturbation theory around free Klein-Gordon fields on curved
space-time is well-defined \cite{BF2}, subject to the same UV
limitations as in flat space-time. Applied to AdS, the interacting
fields may be restricted to the boundary (in the sense of limits of
correlation functions \cite{BBMS}). Thus, canonical bulk perturbation
theory with subsequent restriction provides a new perturbative scheme
around non-canonical free fields.    

Let us return to the Euclidean functional integrals. In the free case,
an identification between $Z^-(f)$ and $\widetilde Z^+(f)$ as in (1.6)
also holds symmetrically between $Z^+(f)$ and $\widetilde Z^-(f)$. The
latter both yield the Gaussian Euclidean field with dimension
$\Delta_-$, and, unless the AdS mass parameter $M$ exceeds the
unitarity bound, the real-time generalized free field with the same
dimension. But in the presence of an interaction,  
the generating 
functionals $Z^+$ seems to be ill defined because a Green function
$\Gamma_+$ decaying faster than $z^{\Delta_+}$ does not exist. 
Eq.\ (1.6), however, suggests to {\em define} $Z^+(f)$ as the
functional Fourier transform of $Z^-(if/c)$. This qualifies and
extends the observation that in the free case the corresponding
connected functionals $\log Z^-$ and $\log Z^+$ are each other's
Legendre transforms \cite{KW}.      
One may doubt, however, that the Fourier transform of the generating
functional respects OS positivity.

On the other hand, we see no a priori obstruction against a field 
theoretical perturbation around the canonical free bulk field 
$\phi^-$ (provided $\Delta_- > \frac{d-2}2$), which then admits a sensible 
restriction $\phi_0^-$. Its generating functional $\widetilde Z^-(f)$, 
however, would have no interpretation as an AdS-CFT functional
integral with $\delta$ function insertion.

We finally notice that we had to chose an (implicit) UV-regularization
in (3.14).  
Our choice seems to be the most natural one, and it gives automatically
the correct normalization required by Ward identities when the scalar
field is coupled to a massless vector field \cite{FMMR,KW}.  
The identification (1.6) suggests that the fulfillment of Ward
identities is another feature which is inherited upon restriction from
the bulk QFT.

\vskip1mm
{\bf Acknowlegment:} We acknowledge partial support of this work by
the Deutsche Forschungsgemeinschaft. We are grateful to D. Buchholz and 
K. Fredenhagen for critical and helpful comments.

\end{document}